# CORTICAL TEMPORAL MISMATCH COMPENSATION IN BIMODAL COCHLEAR IMPLANT USERS: SELECTIVE ATTENTION DECODING AND PUPILLOMETRY STUDY

Hanna Dolhopiatenko and Waldo Nogueira

*Medical University Hannover, Cluster of Excellence 'Hearing4all', Hannover, Germany*

1. ABSTRACT

Bimodal cochlear implant (CI) users combine electrical stimulation from a CI in one ear with acoustic stimulation through either normal hearing or a hearing aid in the opposite ear. While this bimodal stimulation typically improves speech perception, the degree of improvement varies significantly and can sometimes result in interference effects. This variability is associated with the integration of electric and acoustic signals, which can be influenced by several factors, including temporal mismatch between the two sides.

In previous work, we utilized cortical auditory evoked potentials (CAEPs) to estimate the temporal mismatch between the CI stimulation (CIS) side and the acoustic stimulation (AS) side, based on differences in N1 latencies when listening with the CIS alone and the AS alone. Building on this approach, the present study estimates the temporal mismatch and investigates the impact of compensating for this mismatch on speech perception.

Behavioral and objective measures were conducted in bimodal CI users under three bimodal listening conditions: clinical setting, a setting with compensated temporal mismatch between electric and acoustic stimulation and a setting with a large temporal mismatch of 50 ms between electric and acoustic stimulation. The behavioral measure consisted of a speech understanding test. Objective measures included pupillometry, electroencephalography (EEG) based on cortical auditory evoked potentials (CAEPs), EEG based on selective attention decoding including analysis of parietal alpha power. The temporal mismatch between the two listening sides was estimated and compensated based on N1 peak latencies between CAEPs measured with the acoustic side alone and CAEPs measured with the electric side alone.

Despite unchanged speech understanding performance even with a 50 ms temporal mismatch, neural metrics—including CAEPs, selective attention decoding, and alpha power—revealed more pronounced effects. CAEP N1P2 amplitudes were the largest under compensated conditions. Phase-locking value of CAEPs and temporal response function of selective attention also demonstrated an increase compared to clinical settings, however, no significance was observed. Parietal alpha power significantly increased under 50 ms temporal mismatch, indicating unconscious cognitive resource allocation. Pupillometry showed correlation with speech understanding, highlighting its limited sensitivity in the current study.

These findings emphasize that neural metrics are more sensitive than behavioral measures in detecting interaural mismatch effects. A significant enhancement of CAEPs N1P2 amplitude compared to clinical setting was observed. Other neural metrics showed a limited improvement with compensated listening condition, suggesting insufficient compensation solely in temporal domain.

2. INTRODUCTION

Bimodal stimulation combines electric hearing through a cochlear implant (CI) and acoustic hearing through the normal hearing ear or a hearing aid. The use of both stimulations usually results in speech understanding improvements, usually referred to as bimodal benefit (Kong et al., 2005; Dorman et al., 2008; Vermeire and Van de Heyning, 2009; Yoon et al., 2015; Devocht et al., 2017). However, the bimodal benefit exhibits significant variability, and in some instances, even results in interference effects (Ching et al., 2007; Crew et al., 2015; Litovsky et al., 2006; Mok et al., 2006). This variability is associated with the integration of electric and acoustic information (Yoon et al., 2015). This integration might be affected by different factors such as temporal, frequency or level mismatches between both stimulation modalities.

Previous research indicated that achieving effective compensation between the two listening sides in CI users with contralateral acoustic hearing requires addressing multiple dimensions—such as level, timing, and frequency—rather than focusing on a single factor (see Pieper et al., 2022, for a review). However, focusing on correcting temporal delay mismatches between the two sides may provide a promising starting point. This dimension appears to be less influenced by other factors and has the most significant impact on the observed discrepancies. This approach is supported by Wess et al. (2017), where vocoder simulations showed that a substantial latency difference between listening sides greatly diminished the perception of frequency mismatch. Therefore, in the current work, we focus on the temporal mismatch between electric and acoustic stimulation.

Estimating the temporal mismatch between electric and acoustic sides is challenging. One possible method is to measure auditory brainstem responses (ABR). This approach has been used to estimate the temporal difference in wave V latency when listening monaurally to electric and acoustic stimulation (Zirn et al., 2015). Compensation of the temporal mismatch based on the ABR wave V led to an improvement in localization accuracy but not in speech understanding. It is not clear whether compensating for the ABR-level mismatch will result in temporal aligned processing between both modalities at central level. This uncertainty arises from the fact that temporal disparities between electric and acoustic stimulation exist, not only in peripheral timing but also across the brainstem (Polonenko et al, 2015). As the auditory object formation is happening at higher levels of neural processing and believed to be reflected in early cortical activity such as at the N1 peak (Alain and Arnott, 2000, Näätänen and Picton, 1987) we assume that compensation at cortical level could lead to an improvement in speech perception.

In our previous work, we investigated temporal mismatch estimation based on CAEP measurements (Dolhopiatenko et al., 2024). The temporal mismatch in bimodal CI users was compensated based on the difference in N1 latency of CAEPs when listening with acoustic stimulation (AS) alone and cochlear implant stimulation (CIS) alone. The temporal mismatch compensation led to a higher amplitude in the CAEPs recorded with bimodal stimulation. While the phase locking value (PLV) increased for some subjects, the extent was limited, likely due to imprecise temporal compensation. Furthermore, it remains uncertain whether increased CAEP amplitude and PLV would lead to improved speech perception. Building up on the mentioned approach, the current study estimates the individual temporal mismatch as the difference between N1 peak latency when listening with AS alone and with CIS alone.

Furthermore, the effect of temporal mismatch compensation on speech perception is investigated.

Speech perception can be evaluated from various perspectives using both objective and behavioral measures. One common approach to assess speech understanding behaviorally is to measure the percentage of correct recalled words after a participant listens to a given speech material. Such tests require a behavioral response, which may not be feasible for all populations. Furthermore, it is possible that compensating for temporal mismatches, as done in the present study, may have a limited effect on traditional behavioral speech understanding measures. Speech processing relies on envelope cues, which fluctuate between 2 to 5 Hz for syllables and 15 to 30 Hz for phonemes (Elliot & Theunissen, 2009). Furthermore, the precedence effect describes the fusion of the sounds into a single auditory percept within a 30 to 40 ms temporal window (Litovsky, 1999). This effect helps maintain speech intelligibility even in highly reverberant environments and is likely of cortical origin (Miller et al., 2009). However, temporal misalignment still disrupts common onset cues, making auditory grouping more challenging (Bregman, 1990). Auditory grouping is a phenomenon that forms the basis for segregating overlapping speakers in a "cocktail party" scenario (Carlyon, 2004; Bizley and Cohen 2014). Common timing of onsets and offsets of the sound are more likely to form a single auditory object (Darwin and Carlyon 1995; Elhilali et al., 2009) and therefore facilitate speaker segregation.

When a listener successfully follows one of the speakers in a multispeaker scenario, the attended signal modulates cortical neural activity (Ding & Simon 2012, Mesgarani & Chang 2012). Therefore, it is possible to decode the acoustic features of the attended speaker from brain recordings. When a single speaker is presented, the paradigm is usually called neural tracking. When additionally an interference is introduced, the paradigm is usually called selective attention decoding. Previous works have demonstrated that neural tracking and selective attention decoding are related to speech understanding in NH listeners (Iotzov and Parra, 2019; Vanthornhout et al., 2018; Lesenfants et al. 2020; Decruy et al., 2020) and in CI users (Nogueira and Dolhopiatenko, 2022; MacIntyre et al., 2024). Moreover, selective attention decoding reflects not only speech understanding but also the allocation of cognitive resources. Some previous research demonstrated the influence of top-down mechanisms such as attention or prior knowledge on selective attention decoding (Vanthornhaut et al., 2019; Baltzel et al., 2017). When attentional resources are deployed effectively, they enhance the ability to separate target speech from background noise, leading to improved performance (Broadbent, 1958). Furthermore, selective attention functions at the level of objects. If the object formation is limited, for instance due to disrupted synchronization of onsets, there will be a biased competition between objects (Desimone & Dunkan 1995; Kastner & Ungerleider 2000), which might cause a higher cognitive demand. Objective measurements of speech perception could undercover the neural mechanisms underlying speech integration making them a potentially more suitable method for investigating the effects of temporal mismatch compensation on speech neural processing.

In our previous work, we have investigated objective EEG selective attention decoding in bimodal CI users when listening with AS alone, CIS alone and both listening sides (Dolhopiatenko and Nogueira, 2023). We demonstrated that it is possible to decode selective attention in bimodal CI users despite the presence of the CI electrical artifact. Moreover, analysis of selective attention temporal response functions (TRF) revealed a delay in the TRFs obtained with CIS alone compared to those obtained with AS alone. Meanwhile, bimodal TRFs obtained lower amplitude than the TRF estimated from the better ear, which may indicate that the temporal mismatch between individual listening sides led to a suppression of the TRF amplitude when listening with both sides. The current work estimates the temporal mismatch between two listening sides based on N1 latency of CAEPs and investigates the impact of the temporal mismatch correction on behavioral speech understanding and selective attention decoding. Additionally, we included a listening condition with a 50 ms temporal mismatch between listening sides to assess the sensitivity of selective attention measures to temporal mismatch compensation.

Integrating mismatched information from both listening sides may increase cognitive load and consequently require more listening effort (Burg et al, 2022). Meanwhile, individuals who achieve similar speech-in-noise understanding performance may expend different levels of effort (Anderson Gosselin & Gagné, 2010). Listening effort intensifies in challenging environments, such as when processing speech in noisy or reverberant listening conditions (Huang et al., 2022). Therefore, the current work also investigates listening effort with temporally matched and mismatched conditions. Additionally, listening effort was also investigated for a condition with a large temporal mismatch of 50 ms. In a simulation study of Wess (2017), it was shown that interaural delays exceeding 24 ms affect binaural unmasking in single-sided deaf CI users. Additionally, studies on echo effects have shown that when the delay between the target speech signal and its echo exceeds 40 milliseconds, the echo is perceived as "annoying," potentially disrupting cognitive function. (Haas 1951, Grenzebach and Romanus, 2022). Therefore, the 50 ms interaural mismatch condition was included in the current study to investigate the impact of such a large temporal mismatch on the utilized measurements.

One commonly used method for assessing listening effort is measuring pupil dilation. An increase in cognitive task demands consistently leads to greater pupil dilation, making task-evoked pupillary responses a reliable and valid indicator of cognitive processing load (Kahneman & Beatty, 1966, Beatty and Lucerno-Wagoner, 2000, Just et al., 2003, van der Wel and van Steenbergen, 2018). Pupil dilation appears consistent with subjective evaluation of relative difficulty even when intelligibility persists (Koelewijn et al., 2012). Besides pupil dilation, alpha power can also be investigated as a marker of listening effort. Previous work, has demonstrated a relation between alpha activity and self-reported subjective effort in NH listeners (Wöstmann et al., 2015) and in CI users (Dimitrijevic et al., 2019). Moreover, larger pupil responses and larger alpha activity was observed for less intelligible speech (McMahon et al., 2016). Therefore, the current work includes pupil dilation measurements and alpha power to estimate listening effort in bimodal CI users.

The current work estimates the temporal mismatch between two listening sides in bimodal CI users through N1 latency of CAEP measurements. Furthermore, the impact of compensating the temporal mismatch on speech understanding performance, EEG selective attention decoding, alpha power and pupillometry measurements of listening effort is investigated. To determine if the implied measurements are sensitive to the temporal mismatch between listening sides, an additional condition with a large temporal mismatch of 50 ms was included in the current study.

## 3. MATERIAL AND METHODS

**Participants.**

10 CI users with contralateral acoustic hearing participated in the study (5 male and 5 female; mean age: 58.9 years). Three participants had Oticon Medical devices and seven Cochlear Ltd devices. Seven participants used a hearing aid on the contralateral side to the CI. All subjects were native German speakers and had more than 3 years of experience with their implant. Participant's demographics are provided in Table 1. Figure 1 provides the audiogram measured on the AS side in unaided condition. Prior to the start of the experiment, all participants provided written informed consent and the study was carried out in accordance with the Declaration of Helsinki principles, approved by the Ethics Committee of the Hannover Medical School.

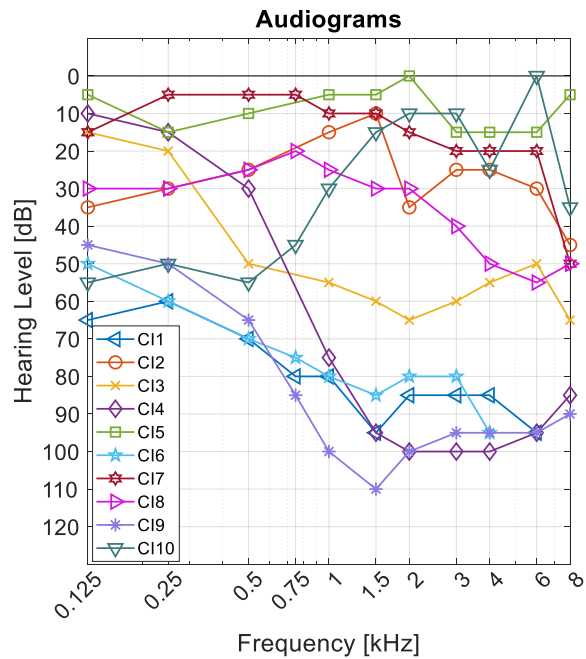

Figure 1. Audiogram measured on the acoustic side in unaided condition.

| ID | Sex | Age (years) | Hearing aid | CI side | CI usage (years) | Stream to attend |
|---|---|---|---|---|---|---|
| CI1 | m | 70 | yes | left | 6.4 | male |
| CI2 | f | 65 | no | right | 6.2 | male |
| CI3 | m | 60 | yes | right | 3.3 | male |
| CI4 | f | 56 | yes | left | 8.1 | female |
| CI5 | m | 52 | no | right | 12.4 | male |
| CI6 | f | 33 | yes | left | 5 | female |
| CI7 | m | 62 | no | left | 6.7 | male |

| CI8 | f | 58 | yes | left | 12.6 | female |
| CI9 | f | 69 | yes | left | 18.3 | female |
| CI10 | m | 64 | yes | left | 11.5 | male |

Table 1. Demographics of participants.

**Data collection**

Participants were invited to take part in two scheduled appointments. In the first session, CAEPs were recorded using EEG. During the second session, behavioral speech understanding was assessed simultaneously with pupillometry. Additionally, EEG recordings were conducted to measure CAEPs and selective attention.
Both experiments were conducted in an electromagnetically and acoustically shielded booth. The luminescence in the cabin was adjusted and kept at 30 lux (Testo SE & Co, Germany). EEG measurements were conducted using a SynAmps RT System with 64 electrodes mounted in a customized, infracerebral electrode cap (Compumedics Neuroscan, Australia). The reference electrode was placed on the nose tip; two additional electrodes were placed on the mastoids. Impedances were controlled and maintained below 15 kOhm. Electrodes with high impedance were excluded from further analysis.

**Experiment 1.**

To estimate the delay between electric and acoustic stimulation, CAEPs were measured when listening with CIS only and AS only. Each subject was instructed to sit relaxed, avoid any movements, and to keep their eyes open in order to minimize physiological artifacts.
A 50 ms broadband noise was used as stimulus. Sound to the AS side was presented through an inserted earphone (E-A-RTONE Gold 3A, 3M, St. Paul, Minneapolis). In case of hearing aid usage, sound was preprocessed through a digital hearing aid implemented in MATLAB (Krüger et al., 2022). For participants implanted with Oticon Medical devices, sound to the CI side was presented via a Bluetooth streamer (Oticon Medical, France). The delay of the Bluetooth streamer was 15 ms and it was compensated. For the participants implanted with Cochlear Ltd devices, stimulation to record CAEPs was delivered through direct stimulation. Direct stimulation was performed through the research interface Nucleus Implant Communicator (NIC, version 4.3.0), a CP910 sound processor and a Programming Pod (all Cochlear Ltd., Australia). Note that direct electric stimulation of the CI allows full control of the delay. Therefore, the CI processing delay of NIC with the Cochlear Ltd implant was measured with the oscilloscope (Pico Technology, UK) and resulted in 11.5 ms. This CI processing delay was considered to compensation the temporal mismatch between CIS and AS. Prior to the experiment, an adjustment was made to ensure equal perceived loudness of the presented stimuli between both listening sides. Initially, the sound presentation level was set independently for the CIS and AS to a loudness level of 7 (loud but comfortable), in a 10-point loudness scale where 1 means inaudible and 10 means extremely loud. Subsequently, sound was presented on both sides, and participants were instructed to compare and, if necessary, modify the presentation level on the CIS side until it matched the perceived loudness on the AS side. Loudness adjustments for the AS side were made in an analog manner using the BabyFace RME (RME, Germany). For the CIS side, presentation level was changed using a Lake People (Lake People electronic GmbH, Germany) audio amplifier for Oticon Medical CI users or by digital amplitude modification of the stimulus for Cochlear Ltd CI users. Stimulus was presented 100 times per each condition with a presentation rate of 1 second. The EEG data was recorded with 20 kHz sampling rate.

The recorded EEG data were processed using the EEGLAB MATLAB toolbox (Delorme & Makeig, 2004). To eliminate electrical artifacts from CI, independent component analysis (ICA) based on second-order blind identification (SOBI; Belouchrani et al., 1993) was applied to the data. Following the SOBI procedure, topoplots and waveforms of the components were manually inspected. Components exhibiting activation at the CI site and displaying clear electrical artifacts in the temporal domain were removed from the EEG data. With a sampling rate of 20 kHz, the electrical artifacts were easily identifiable in both the raw EEG data and the temporal domain of the computed ICA components. The cleaned EEG data were then epoched within a time interval of -200 ms to 1000 ms. The signal was filtered between 1 and 15 Hz and re-referenced to the mean of the two mastoid electrodes. The amplitude of the N1 peak was estimated for responses during listening conditions with only AS and only CIS. The difference in N1 latency between these two responses was interpreted as an individual temporal mismatch between the two listening modalities at cortical level.

**Experiment 2.**

Based on the results from Experiment 1, the material for experiment 2 was prepared and the following three listening conditions were generated:
- Bimodal clinical (Clin) listening condition with the clinical settings of each participant without any temporal mismatch compensation;
- Bimodal compensated (Comp) listening condition with temporal mismatch compensation. The presented material was temporally delayed on the AS side or CIS side based on the estimated N1 latency difference between AS only and CIS only. The delay in the presented material was achieved by inserting zero values into the sound channel of interest.
- Bimodal 50 ms (50ms) listening condition where the AS listening side was delayed 50 ms with respect to the compensated condition. This condition was measured to investigate the impact of a large temporal mismatch on speech perception, listening effort and selective attention.

In case of hearing aid use on the AS, all speech material was preprocessed with a digital hearing aid implemented in MATLAB based on the hearing thresholds of the participant. This was conducted to reduce variability caused by hearing aid processing delay. The digital hearing aid applies the half-gain rule for amplification. All speech material used in Experiment 2 was presented to the AS side using an inserted earphone (E-A-RTONE Gold 3A, 3M, St. Paul, Minneapolis). To the CIS side for Oticon Medical CI users, sound was presented using a Bluetooth Streamer, the delay of which was compensated in the presented material. For Cochlear Ltd CI users the CP910 research sound processor was programmed with the participant's actual fitting map and speech material was presented using direct cable.

*Behavioral Test.* The German Hochmair-Schulz-Moser sentence test (HSM test) (Hochmair-Desoyer et al., 1997) was used to assess speech understanding behaviorally. Each list consists of 20 semantically structured sentences, uttered by a male or by a female talker. Target and interference speech streams were co-located and presented at 0 dB signal-to-interference ratio (SIR) between the target (male/female) and the interference (female/male) speech stream. Subjects were instructed to attend to the target talker and to repeat all words after each sentence. The speech stream to be attended was randomized within subjects and is indicated in Table 1. The attended speech stream was kept the same through the whole experiment. Two sentence lists were presented per listening condition. The speech material was presented through Presentation® software (Version 23.0, Neurobehavioral Systems, Inc., Berkeley, CA) using diotic presentation i.e. the same speech material presented to both ears. The presentation order of listening conditions was randomized across subjects. The speech understanding performance score was calculated in percentage of correct recalled words per listening mode. Afterwards, the percentage of correct scores was transformed using a rationalized arcsine transform (RAU units) (Studebaker, 1985).

*Pupillometry.* Pupil dilation was measured using the PupilLabs Core eye tracking system (PupilLabs, Germany). The pupil measurement was conducted together with the behavioral speech understanding test, therefore, the sentence presentation was adjusted accordingly (Figure 2). First, a baseline of 1 second before starting the sentence was recorded. Afterwards, a sentence was presented followed by a retention period of 3 sec where the word "WARTEN" (eng. "WAIT") was presented on the screen. Afterwards, the word "ANTWORTEN" (eng. "ANSWER") appeared on the screen which was a cue for the participant to recall the heard words. After completing the verbal response a keyboard button was pressed by the experimenter. Afterwards, a 5 sec period of silence was followed to let the pupil get back to the baseline. This was completed for each sentence with a break between each list of the presented sentences. At each step, a trigger was sent to the pupil recording app using Lab Streaming Layer (LSL)(1).

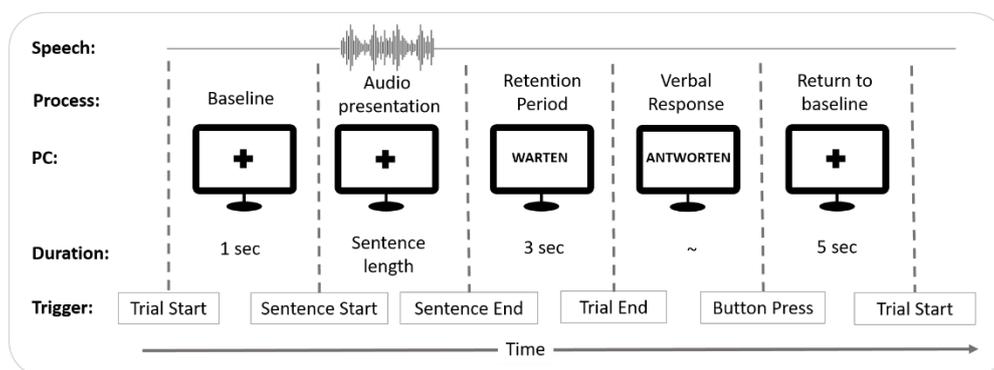

Figure 2. Demonstration of each trial for the speech understanding test adjusted to measure pupillometry.

The pupil data was imported and analyzed using MATLAB. The imported pupil data was segmented in trials corresponding to each sentence by using recorded timestamps. Eye blinks were detected using the median absolute deviation method described by Kret and Sjak-Shie

(2018). Additionally, data recorded 35 ms before and 100 ms after the eye blink was removed. In a case of 30% of missing data within one trial, the trial was discarded. The baseline from 0 to 1 s was calculated and subtracted from each trial. If the eye blink was detected during the baseline, the corresponding trial was discarded. After preprocessing, each sentence trace was aligned to the onset of the sentence and the average trace per participant and per listening condition was calculated. Max pupil dilation was defined as the max of the mean curve relative to baseline within a time window ranging from the trigger 'Sentence Start' till the trigger 'Trial End' (Figure 2) i.e. from the sentence presentation till the start of the verbal response.

*EEG test.* During the electrophysiological test, CAEPs and selective attention were measured. CAEPs were measured following the same procedure as in Experiment 1 and were measured for the Clin, Comp and 50ms listening conditions. For the selective attention decoding paradigm, HSM sentence lists with a male and a female talker at 0 dB SIR were used. The speech stream to be attended was kept the same as in the behavioral part of the experiment. The speech material was presented using the same set up as in the behavioral part of Experiment 2. For each listening condition 8 lists were presented, resulting in approximately 6 min of stimulation per listening condition. To extend the training dataset for the linear decoder, additional speech material consisting of two audio story books were used. The story books included two German narrations ("A drama in the air" by Jules Verne, narrated by a male speaker and "Two brothers" by the Grimm brothers, narrated by a female speaker) at 0 dB SIR. In total, 36 min of story (12 min per listening mode) were presented. To ensure the continuous engagement of the CI user when listening to the corresponding speaker, questions to the context of the presented speech material were asked every 2 min. The presented speech material was randomized across listening conditions to avoid the influence of the material. EEG data for selective attention was recorded with a sampling rate of 1000 Hz.

EEG data was processed offline in MATLAB and the EEGLAB toolbox (Delorme& Makeig, 2004). SOBI artifact rejection was applied separately to CAEP EEG data and to selective attention EEG data to suppress physiological and CI electrical artifact. The time course and topolot of each component was analyzed in order to eliminate CI electrical artifact from the data.

For CAEPs, the amplitude and the latency of the N1P2 peak was estimated for each participant and each condition. Moreover, the PLV was analyzed for each listening condition. The phase spectrogram was obtained by calculating the phase angle of the short time Fourier transform with a Hamming window of 400 ms and 20 ms overlap using MATLAB's spectrogram function. The PLV was calculated as the absolute value of a complex mean phase spectrogram. To investigate a change in phase synchrony relative to the baseline, a baseline normalization using the interval of 200 ms before stimulus onset was applied. The maximum PLV was estimated from the interval between 50 and 200 ms after stimulus onset and the frequency range from 1 to 15 Hz.

For selective attention decoding the EEG data was split into trials corresponding with the duration of each sentence list and 1 min segments of the story. Next, the digital signal was band-pass filtered for frequencies 2–8 Hz and downsampled to 64 Hz. The envelopes of the original attended and unattended speech streams were extracted through the Hilbert transform. The envelopes were filtered with a low-pass filter having cut-off frequency of 8 Hz and downsampled to 64 Hz. Selective attention was analyzed using the forward model approach (Crosse et al., 2016). By applying the forward model, the TRF is obtained. TRF reassembles the N1P2 complex of CAEP, therefore, the general morphology and amplitude and latency of the TRF N1 and TRF P2 peaks was analyzed for each individual. The model was estimated at 500 ms time lag window. The time lag performs a time shift of the EEG signal that reproduces the physiological delay between the audio presentation and its processing up to the cortex (O'Sullivan et al., 2015). The regularization parameter λ, which is applied to the least mean square solution to avoid overfitting, is set to 1000 to maximize the peak amplitudes of the TRF. A classical leave-one-out cross-validation approach was used to train and test the decoder. HSM lists and the story were used to train the decoder. More details on the procedure can be found in Nogueira et al. (2020).

Alpha power was calculated for the selective attention EEG data. To achieve this, clean data were averaged across trials. The Welch method with window of 512 samples and 50% was applied to estimate power across different frequencies. Alpha power was analyzed within the frequency band of 8 to 13 Hz and was expressed as a dB change relative to the baseline, using the following equation:

$$Alpha\ Power = 10 \cdot log_{10}(Welsch_{Main}/Welsch_{Base}). \quad (1)$$

Here, $Welsch_{Base}$ represents the Welch power estimated during the baseline period from -1 to -0.05 seconds before the stimulus onset, while $Welsch_{Main}$ refers to the power calculated over the trial duration from 0 to 7 s. Only the first 7 s from each trial were used because alpha power is higher at the start of the trial and afterwards decreases which is associated with a decline in attentional controls in a speech-in-noise task (Wöstmann et al., 2015). The dB change in alpha power for each participant was averaged across 13 electrodes located in the parietal region.

## 4. RESULTS

**Experiment 1**

***CAEPs.*** Figure 3 demonstrates mean CAEPs measured during Experiment 1. The responses were measured when listening with AS only and CIS only. For each participant, the maximum N1 amplitude and its corresponding N1 latency were determined for both listening conditions. The temporal mismatch between the two listening sides was quantified as the difference in the N1 latency with each condition and demonstrated in Table 2.

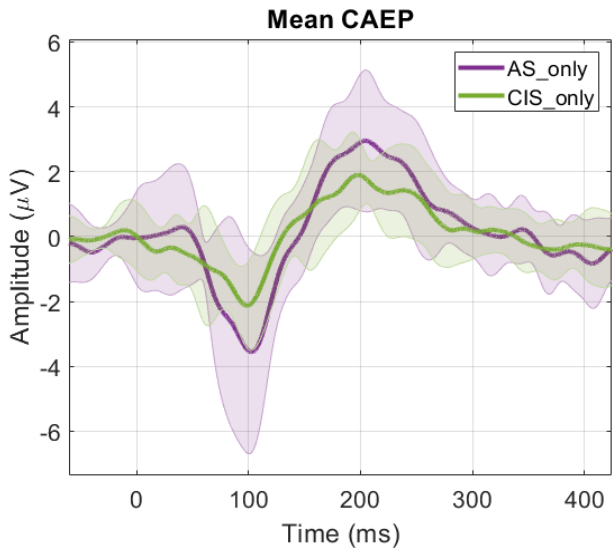

| ID | Mismatch (AS_only-CIS_only) |
|---|---|
| CI1 | -10 ms |
| CI2 | -1 ms |
| CI3 | 12 ms |
| CI4 | -14 ms |
| CI5 | 20 ms |
| CI6 | 4 ms |
| CI7 | 2 ms |
| CI8 | 5 ms |
| CI9 | -14 ms |
| CI10 | 11 ms |
| mean | 1.5 ms |

Figure 3. Mean cortical auditory evoked potentials (CAEPs) across subjects for listening conditions acoustic stimulation (AS) only and cochlea implant stimulation (CIS) only. The thick line represents the mean and the shaded area represents the standard deviation value.

Table 2. Temporal mismatch between acoustic stimulation (AS) only and cochlear implant stimulation (CIS) only estimated by subtracting the CIS_only N1 latency from the AS_only N1 latency for each subject. Positive and negative value of estimated mismatch indicate the precedence of the AS side or CIS side, respectively.

**Experiment 2.**

*Behavioral test.* Figure 4 presents the results of the behavioral HSM speech-on-speech understanding test as percentage of correct recalled words (in RAU units). Results are presented for each individual when listening with the Clin, Comp and 50ms listening conditions. No impact of listening condition on speech understanding can be observed.

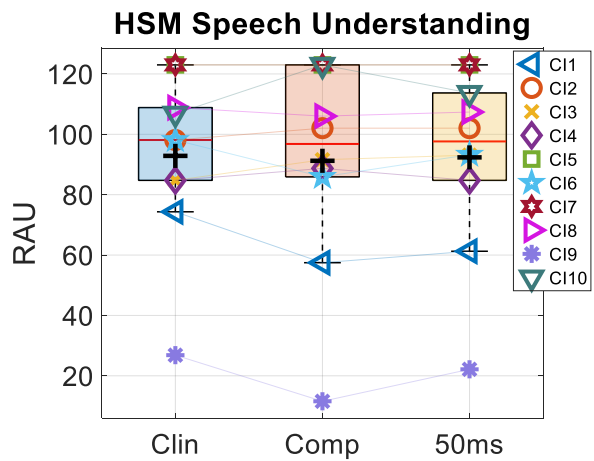

Figure 4. Behavioral HSM speech understanding performance across three listening conditions: clinical (Clin), compensated (Comp) and 50 ms interaural mismatch (50ms). The results are presented rationalized arcsine transform (RAU units)

*Pupilometry.* Figure 5 (left) displays the mean pupil traces averaged across subjects under three listening conditions: Clin, Comp, and 50ms. On the right, Figure 5 presents a boxplot with the individual mean pupil dilation of each participant's trace calculated across trials (sentences) for each condition. The max pupil dilation was calculated relative to the baseline and estimated as the maximum at a time interval starting with the sentence presentation up to the end of the retention period before verbal response. No effect of listening condition on pupil dilation was observed. A repeated measures ANOVA (rmANOVA) revealed no effect of condition on pupil dilation ($F(2,18) = 0.883$; $p = 0.431$). A post-hoc also did not show significance for any of the pairs.

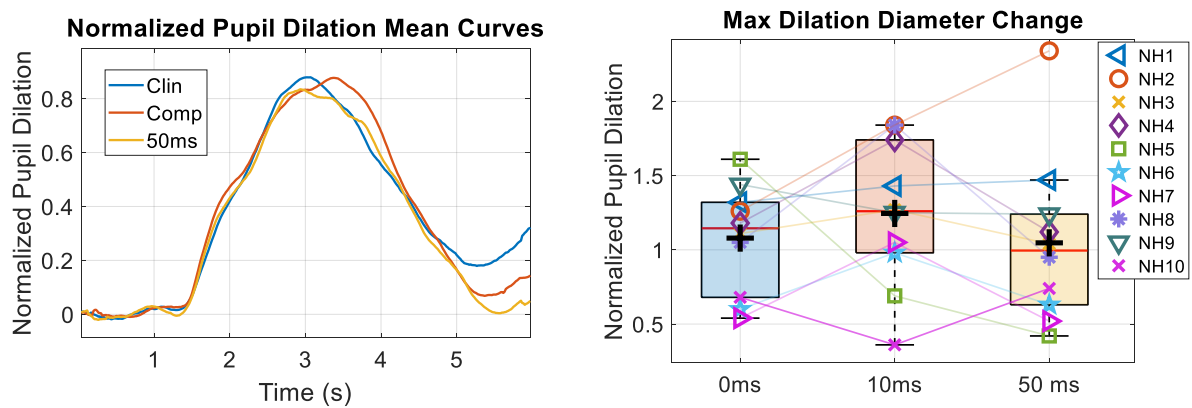

Figure 5. Left: Mean pupillometry curves averaged across subjects for three listening conditions: clinical (Clin), compensated (Comp) and 50 ms interaural mismatch (50ms). (b) Mean pupil dilation estimated for each subject during the sentence presentation.

A Spearman correlation between HSM speech understanding RAU units and mean pupil dilation for each condition was investigated (Figure 6). A linear trend between both measures is observed, however, the result was only significant for the dataset when listening with 50ms interaural mismatch ($p = 0.01$), probably due to the higher span in HSM speech understanding performance scores with this condition.

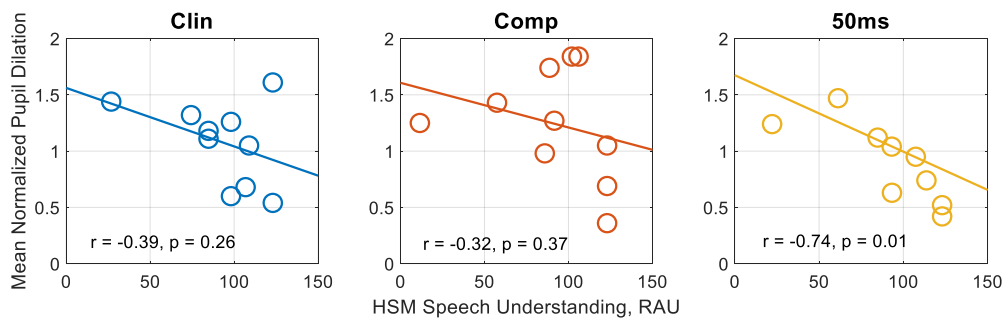

Figure 6. Spearman Correlation between HSM speech understanding RAU units and mean normalized pupil dilation for three listening conditions: clinical (Clin), compensated (Comp) and 50 ms interaural mismatch (50ms).

*CAEPs.* Figure 7 (left) presents the mean CAEPs across subjects for the Clin, Comp and 50ms listening conditions. Similar N1 and P2 peak morphologies, amplitudes and latencies for the Clin and Comp conditions were observed, while the Comp condition presented slightly larger P2 amplitude. For the condition 50ms a reduction in N1 amplitude and a delay in P2 latency can be visually observed. Additionally, the 50ms condition reveals two distinct N1 peaks, which can be attributed to the elicitation of separate responses from each listening side. To investigate it further, individual N1 and P2 peak amplitudes and latencies were estimated. Afterwards, the N1P2 amplitude was calculated as the sum of the absolute values of the N1 amplitude and the P2 amplitude. The estimated N1P2 amplitude for each individual is demonstrated on Figure 7 (right). A rmANOVA was applied to investigate the impact of listening condition on the N1P2 amplitude and revealed a significant effect ($F(2,18)=20.456; p < 0.001$). A post-hoc one-tailed t-test with Bonferoni-Holm correction was applied and revealed a significant difference between N1P2 amplitude for the pair Comp and 50ms ($p < 0.001$), the pair Clin and 50ms ($p = 0.002$) and the pair Clin and Comp ($p = 0.025$).

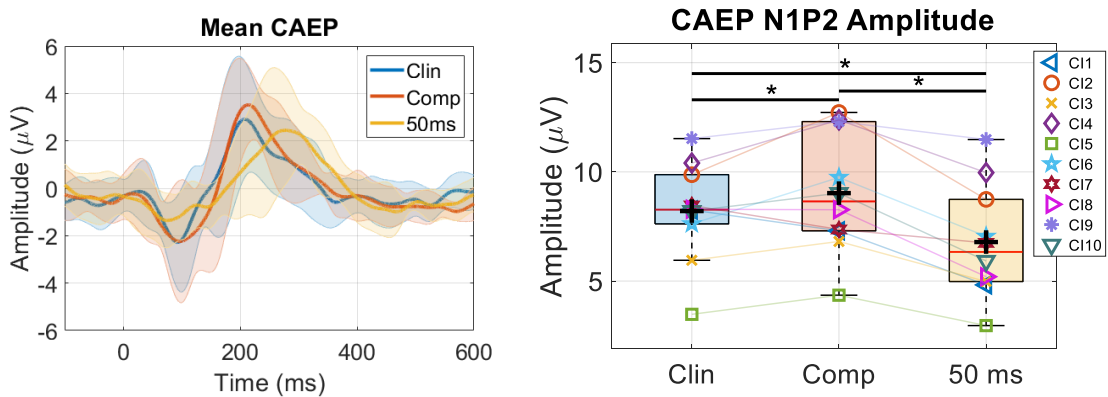

Figure 7. Left: Cortical auditory evoked potentials (CAEPs) averaged across subjects for clinical (Clin), compensated (Comp) and 50 ms interaural mismatch (50ms) listening conditions. The thick line represents the mean and shaded area represents standard deviation. Right: N1P2 amplitude for each individual across three listening conditions. The black cross represents the mean across individuals. The black line with asterisk indicates significances for the pair of conditions revealed through a t-test with Bonferroni-Holm correction.

The PLV for each condition was calculated for each subject across listening conditions. Figure 8 presents the average time-frequency representations of the PLV across participants for three listening conditions. Visually, an increase of PLV for the Comp listening condition compared to Clin and 50ms can be observed. Figure 9 shows the max PLV for each individual. A rmANOVA revealed a significant impact of listening condition on PLV ($F(2, 18)=4.259; p = 0.031$). A t-test using Bonferroni-Holm correction showed significance for the pair Comp and 50ms ($p = 0.010$). Other pair of comparisons were not significant: Clin and 50ms ($p = 0.073$), Clin and Comp ($p = 0.098$).

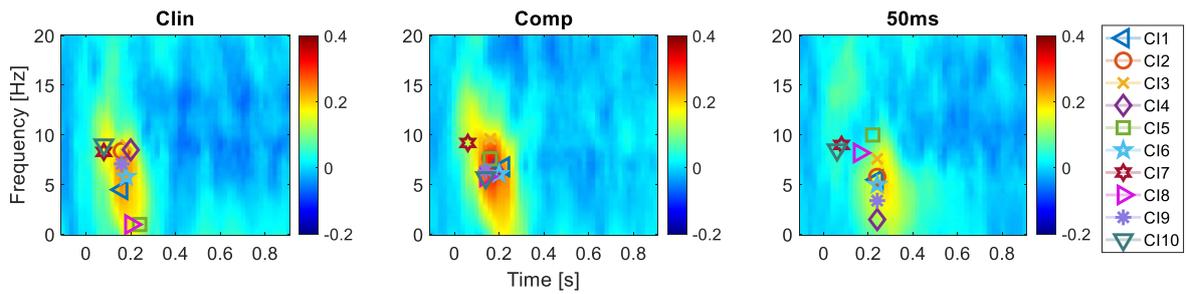

Figure 8. Mean phase locking value (PLV) time-frequency representation for different listening conditions. (clinical: Clin, compensated: Comp, 50 ms interaural mismatch: 50ms). The markers indicate the maximum PLV for each individual.

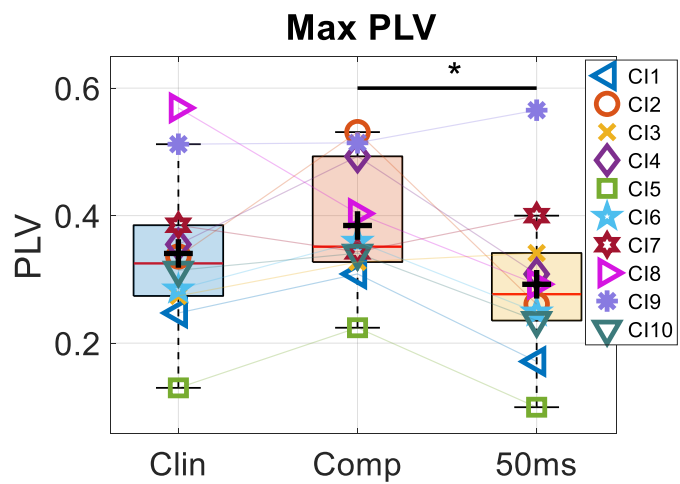

Figure 9. Boxplots representing the maximum phase locking value (max PLV) for each individual. The markers indicate the results for each individual subject and for each listening condition (clinical: Clin, compensated: comp, 50 ms interaural mismatch: 50ms). The black line with asterisk indicates significances for the pair of conditions revealed through a t-test with Bonferroni-Holm correction.

*Selective Attention Decoding.* Figure 10 presents the weights for the attended and unattended TRFs averaged across all participants per each listening condition. It can be observed that the Clin and Comp attended TRFs have a more robust morphology in comparison to the 50ms TRF. Moreover, for the 50ms listening condition the attended and unattended TRFs are not visibly different at the vicinity of the TRF N1 peak or the TRF P2 peak compared to the Clin and Comp conditions.

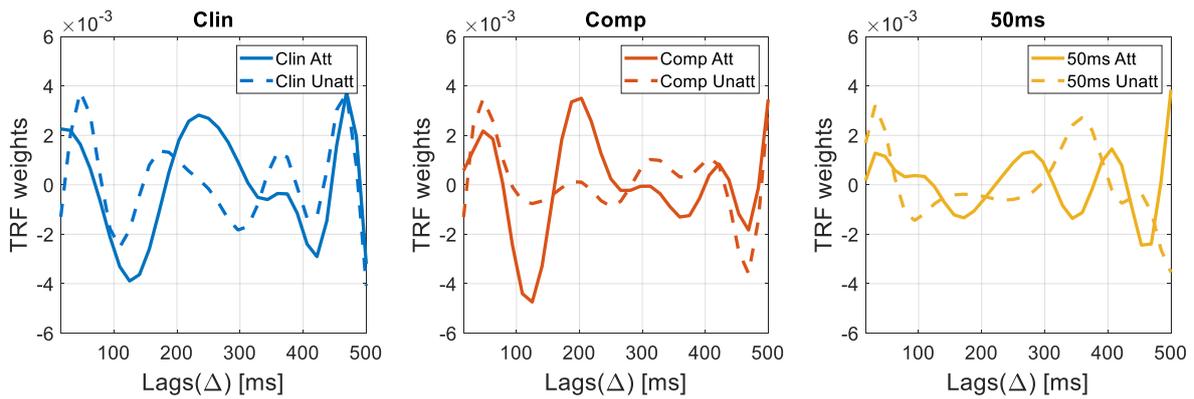

Figure 10. Attended (solid) and unattended (dashed) TRFs averaged across participants per each listening condition: clinical (Clin), compensated (Comp) and 50 ms interaural mismatch (50ms).

Figure 11 (left) presents the averaged attended TRF for each listening condition for a better visualization of the differences across listening conditions. For each individual, a maximum attended TRF N1 amplitude and attended TRF P2 amplitude were estimated. Figure 11 (right) shows the topoplot for the TRF N1 peak and TRF P2 peak averaged across participants for the three listening conditions.

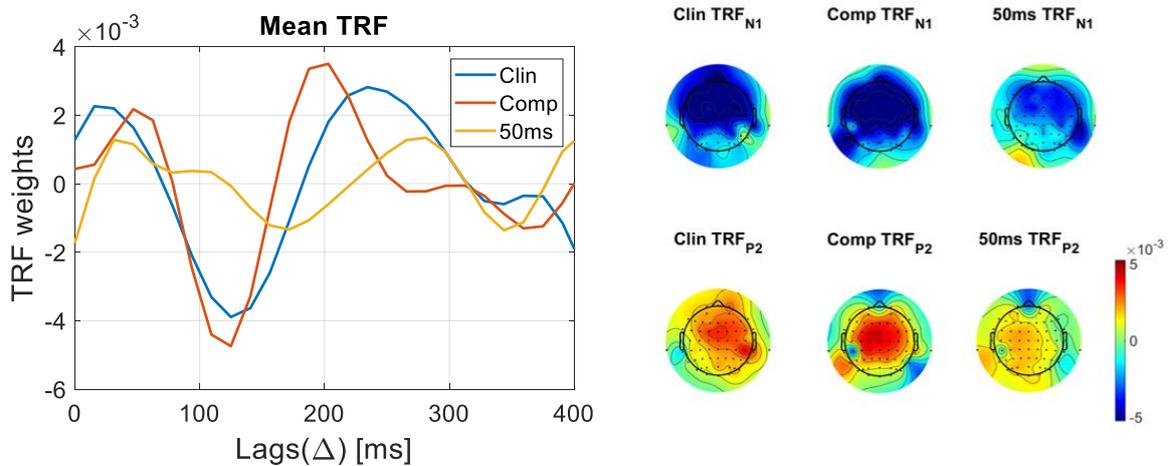

Figure 11. Left: Mean attended TRF weights averaged across subjects for the clinical (Clin), compensated (Comp) and 50 ms interaural mismatch (50ms) listening conditions. Right: Topoplots averaged across subjects at the TRF N1 and P2 peak latencies.

Figure 12 presents the individually estimated TRF N1P2 amplitudes for each listening condition. A rmANOVA revealed a significant impact of listening condition on the TRF N1P2 amplitude ($F(2,18)=5.420$; $p = 0.014$). A post-hoc t-test was applied and a significant difference was found for the pair Comp and 50ms ($p = 0.008$) and the pair Clin and 50ms ($p = 0.018$). No significant difference between Clin and Comp was observed ($p = 0.192$).

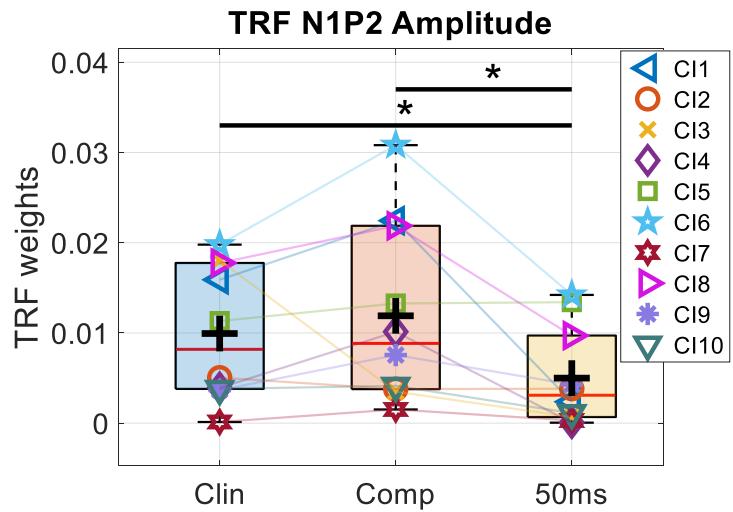

Figure 12. Attended TRF N1P2 amplitude estimated per each individual across three listening conditions: clinical (Clin), compensated (Comp) and 50 ms interaural mismatch (50ms). Black crosses correspond to the mean value. The thick line with the asterisk indicates a significance.

Alpha power was calculated from selective attention EEG data. The alpha power for each condition and each individual was calculated as the dB change relative to the baseline. Figure 13 presents the topolplots for the alpha power averaged across subjects per each listening condition. A typical activation in the parietal part across all conditions can be observed. Following this analysis, the alpha power dB change was averaged for each participant across 13 parietal electrodes (Figure 14).

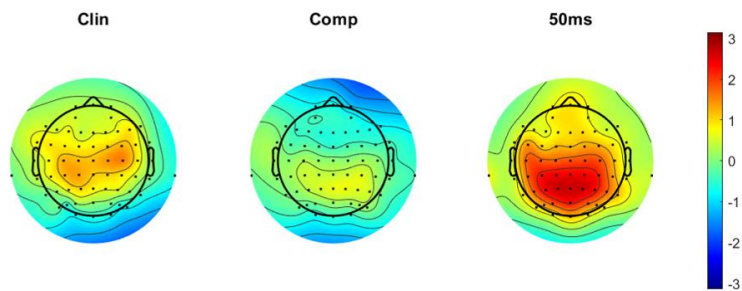

Figure 13. Topoplot of alpha power dB change distribution averaged across subjects for the three listening conditions clinical (Clin), compensated (Comp) and 50 ms interaural mismatch (50ms). Alpha power is demonstrated as difference relatively to the baseline calculated in dB.

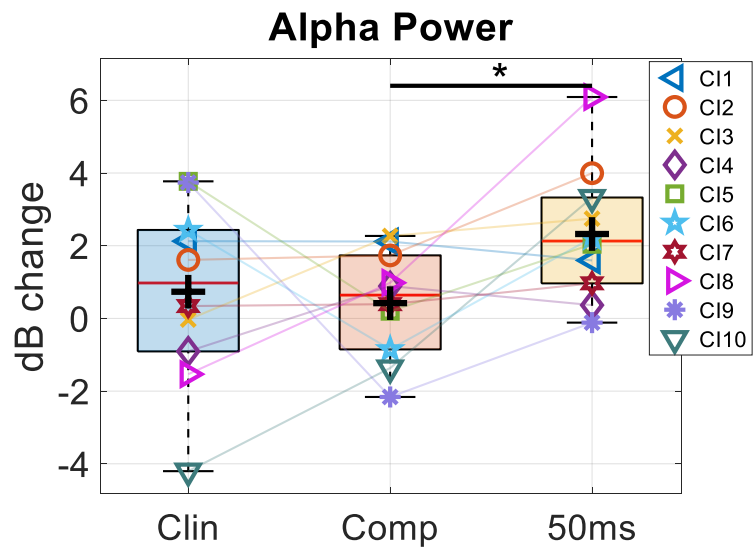

Figure 14. Individual alpha power difference relative to the baseline in dB across three listening conditions and averaged across parietal electrodes. Black cross represents mean value. Black line with asterisk indicates significance.

A rmANOVA revealed no significant impact of listening condition on parietal alpha power ($F(2,18)=2.354$; $p = 0.124$). However, a post-hoc t-test pairwise comparison revealed a significant difference for the pair of Comp and 50ms ($p = 0.007$). The pairs Clin vs 50ms ($p = 0.104$) and Clin vs Comp ($p = 0.373$) were not significantly different.

## 5. DISCUSSION

Bimodal stimulation typically leads to improvements in speech perception, but the extent of this improvement varies significantly across individuals, with some bimodal CI users obtaining interference effects. One potential factor contributing to this variability is the temporal mismatch between the two ears. In this study, we estimated the temporal mismatch for each individual through cortical measures and explored the effects of correcting this mismatch. The impact of the correction was evaluated through multiple measures, including behavioral speech understanding, listening effort assessed via pupillometry and alpha power, CAEPs and selective attention decoding. The main goal was to clarify how temporal alignment at cortical level influences auditory performance and cognitive load in bimodal listeners.

*Estimation of the temporal mismatch*

To estimate cortical temporal mismatch, CAEPs were measured when listening with AS only and CIS only. The mean amplitude of the N1 peak of CAEPs for each individual was $-4.2\pm3.1$ µV for AS only and $-2.6\pm1.5$ µV for CIS only. These values are in agreement with previous studies showing a more negative N1 amplitude response on the AS side compared to the CIS side (Voola et al., 2023; Dolhopiatenko et al., 2024). The average latency of the N1 peak resulted in $97.1\pm10.4$ ms and in $97.9\pm22$ ms for AS only and CIS only responses, respectively. These findings are also consistent with previous studies that reported no significant difference in averaged auditory responses between the two listening modalities in CI users with contralateral acoustic hearing (Sasaki et al., 2009; Wedekind et al., 2020). On average, the observed mismatch was 1.5±10.7 ms, indicating a small latency difference between the two conditions. However, when examining individual cases, there was substantial variability, with temporal mismatches ranging from -14 ms to 20 ms. Negative values indicate precedence of the AS side, while positive values indicate that the CIS side response preceded the AS side. Therefore, this variability highlights that while the average latency difference is small, individual differences in temporal alignment between acoustic and CI hearing can be quite pronounced. In the current study, the individually estimated temporal mismatch was compensated and its impact on behavioral and objective test was investigated.

*Effect of temporal mismatch across listening sides on behaviorally measured speech understanding*

Compensating for temporal mismatch did not significantly impact behavioral speech understanding performance. The large temporal mismatch of 50 ms also did not impact speech understanding. This result differs from the findings of a vocoder simulation study which reported a substantial drop in binaural unmasking task when the interaural delay exceeded 24 ms (Wess et al., 2017). However, the mentioned study used a binaural paradigm in NH listeners who had a preserved binaural system, the preservation of which is still under debate in bimodal CI users (Dieudonne et al., 2020). Moreover, in NH subjects the auditory information arriving over a large temporal window might be fused in a single auditory object (Litovsky, 1999). Studies investigating the perception of echo and reverberation showed that it is possible to integrate the main sound with a delayed copy ranging from 25 to 50 ms (Nábělek & Robinette, 1978; Warzybok et al., 2013). This result could explain why the large temporal mismatch condition did not affect behavioral speech understanding in the current study. However, it has been shown that an uncommon onset in the speech signal can disrupt auditory grouping (Bregman 1990; Drawin and Carlyon 1995; Elhilali et l., 2009), a process that contributes to segregation of speakers and selective attention (Carlyon 2004). Moreover, studies on echo effects showed that adding the copy of the sound to the target sound with delays exceeding 40 ms are perceived as an annoying "echo" (Haas 1951) which may cause an increase in listening effort similar to the increase observed when listening in a reverberant environment (Huang et al. 2022). Therefore, our hypothesis was that the temporal mismatch across ears may affect neural measurements and listening effort, even if no effect on speech understanding is observed.

Interestingly, at the end of the experiment, participants were asked if they noticed any perceptual difference between the listening conditions, despite not being informed about the specific condition they were listening at any given time. None of the participants reported perceiving any differences, a finding supported by the consistent subjective ratings on listening effort provided by each participant across all conditions (results are presented in Appendix 1). The behavioral test was additionally conducted in seven normal hearing listeners (NH) to investigate the impact of temporal mismatch on speech understanding. The NH group listened to sentences when no temporal mismatch was present (0ms), with a small temporal mismatch (10ms) and a large temporal mismatch of 50 ms (50ms). The listening condition did not affect speech understanding and all NH listeners obtained nearly 100% word-scores across the three listening conditions (results are not presented in the current paper). However, all NH participants reported perceivable differences between listening conditions without being aware that the listening material was modified. This is different to the reports provided by bimodal CI users, who did not perceive any difference between listening conditions. This can be partially explained by the abnormal broad fusion observed in bimodal CI users, assuming that bimodal CI users are integrating the information from both listening sides across a wider temporal window compared to NH listeners (Reiss et al., 2014). More research on binaural/bimodal fusion across different delays in NH listeners and bimodal CI users is recommended.

*Effect of temporal mismatch across listening sides on listening effort measured through pupil dilation*

Having temporally mismatched auditory information may cause higher cognitive demands resulting in elevated listening effort. Winn et al. (2015) using vocoded speech did not observe a substantial effect of degraded speech on speech intelligibility, instead a significant effect on pupil dilation was observed. Furthermore, a study on bilateral CI users demonstrated greater pupil dilation when listening with both implants compared to using only the better one. This suggests that the additional effort required to integrate auditory information from both ears contributes to an increased cognitive load (Burg et al., 2022). Moreover, listening effort intensifies when processing speech in noisy settings or reverberation (Huang et al., 2022). Therefore, the current study hypothesized higher listening effort when a temporal mismatch across ears is present. For this purpose, the current study included pupil dilation measurements during behavioral tasks to assess listening effort when speech understanding remains the same. However, no significant effect of listening condition on pupil dilation was observed, showing that pupil measure was not sensitive enough to capture temporal mismatch compensation effects. One explanation for this could be the absence of any perceptual difference between the listening conditions. Since pupil dilation reflects the mental effort or "will" to perform a task (Pichora-Fuller et al., 2016), it is plausible that when participants do not perceive a difference between conditions, the cognitive load remains similar, leading to comparable pupil dilation across conditions. This idea is partially supported by the observed trend in correlation between speech understanding performance and pupil dilation reported in the current work. Note, that significant Spearman correlation ($r = 0.74$; $p = 0.01$) was observed only for the condition of 50 ms, due to the higher variability in speech understanding results with this condition. On the other hand, the speech-on-speech task may have been overly challenging for bimodal CI users, as informational masking requires more listening effort compared to a non-intelligible noise (Koelewijn et al., 2012), possibly causing participants to disengage or give up on the task (for a review see Winn et al., 2018). This could also explain the lack of differences in pupil responses across listening conditions

*Effect of temporal mismatch across listening sides on EEG measures*

An effect of temporal mismatch compensation on CAEPs and selective attention decoding was investigated. Both measures were recorded when listening with and without compensated temporal mismatch across sides and with the 50 ms interaural mismatch condition.

For CAEPs, the N1P2 amplitude was estimated for each individual. A significant increase in N1P2 amplitude with the Comp listening condition compared to the Clin and 50ms listening conditions were observed. An analysis of PLV also revealed an increase, however, this effect was significant only for the pair Comp and 50ms. These results support our previous study, which demonstrated that temporal compensation at cortical level leads to enhanced CAEP amplitude and PLV (Dolhopiatenko et al., 2024). Similarly to our previous work, an effect of temporal compensation on PLV in bimodal CI users is limited compared to NH listeners, which can be also explained by the lack of binaural processing in bimodal CI users (Dieudonne et al., 2020). Nevertheless, the increase of PLV value with the Comp condition compared to the Clin condition was observed in 8 out of 10 participants.

We hypothesized that more synchronous arrival of the sounds to the cortex from both listening sides facilitates selective attention decoding. To investigate this hypothesis, selective attention decoding was analyzed in terms of TRFs. The morphology of the TRF and the difference between the attended and unattended TRFs were noticeably disrupted when a large temporal mismatch of 50 ms was introduced. This finding highlights that objective measures, such as selective attention decoding, are more sensitive in detecting compensation effects than the behavioral measures employed. To further assess neural responses, peaks resembling the N1 and P2 components of CAEPs were identified for each individual and were termed N1P2_TRF. A significant reduction in N1P2 amplitude was observed for the 50ms listening condition compared to the Comp ($p = 0.008$) and Clin ($p = 0.018$) listening conditions. To the best of our knowledge, this is the first study showing effect of a large temporal mismatch on selective attention decoding in CI users. Previous studies have shown that the tracking of speech envelope from neural recordings correlates with speech intelligibility (Iotzov and Parra, 2019; Vanthornhout et al., 2018; Lesenfants et al., 2020; Decruy et al., 2020; Nogueira and Dolhopiatenko, 2022; MacIntyre et al., 2024). Additionally, increased TRF peak amplitudes have been reported in binaural unmasking paradigms (Dieudonné et al., 2024). In the current study, despite preserved speech understanding, the current work demonstrated a significant reduction in N1P2_TRF amplitude with a large temporal mismatch between listening sides. This result suggests that selective attention decoding reflects neural effects that extend beyond speech intelligibility and may serves as a more suitable measure for investigating the impact of temporal mismatches. Furthermore, the N1P2_TRF amplitude was, on average, higher in the Comp listening condition compared to the Clin condition. Although an increase in N1P2_TRF amplitude was observed in 9 out of 10 CI users, the increase was not statistically significant. Therefore, the question arises whether compensating solely in the temporal domain is sufficient to observe significantly increased TRFs. Future research should explore whether compensatory mechanisms across domains other than temporal processing could further enhance selective attention decoding.

*Effect of temporal mismatch across listening sides on listening effort measured through alpha power*

Besides TRF amplitudes of selective attention decoding, parietal alpha power during the selective attention task was analyzed. A visible increase in alpha power when listening with the 50ms interaural delay condition was observed. Increased alpha power is associated with increased difficulty and subjective listening effort (Dimitrijevic et al., 2019; Wöstmann et al., 2015; Obleser et al., 2012), we hypothesized that when a large temporal mismatch between two listening sides is present, CI users employ more cognitive resources without conscious awareness. A significant difference in alpha power was found between the Comp and the 50ms temporal mismatch conditions (p=0.007) suggesting less effort when temporal mismatch is compensated, even though speech understanding remains the same. No significant difference in alpha power was observed between the Comp and Clin listening conditions. This lack of a distinct effect may be explained by the fact that the bimodal users used the Clin condition in their clinical map and therefore were adapted it. Such adaptation has been documented in previous studies, particularly regarding pitch perception in bimodal CI users (Reiss et al., 2015). Over time, users may adjust to mismatches, minimizing the effort required for auditory processing under less extreme conditions. However, another possibility is that effective compensation for temporal mismatch may need to address multiple auditory domains—beyond just temporal alignment—to reduce listening effort in this population. For example, integrating spectral and temporal information in a cohesive manner could alleviate cognitive load. Future research should explore whether compensating not solely in temporal domain could further reduce cognitive load for CI users.

The current study shows that pupillometry and alpha power revealed different results despite both being measures of listening effort. A relationship between pupil dilation and alpha power becomes less evident under more challenging listening conditions (McMahon et al., 2016). Given that our study employed a speech-on-speech test, which is particularly challenging for CI users, this could explain the absence of a relationship between these two measures. However, a linear correlation was found between pupillometry and speech understanding with 50ms listening condition, whereas alpha power did not show any correlation with speech understanding. This aligns with findings by Miles et al. (2017), which reported that pupil dilation was linked to intelligibility scores, unlike EEG alpha power, suggesting that distinct cognitive mechanisms may be contributing to each of these measures. Since pupillometry correlated more closely with speech understanding in the current study, we propose that it may partly reflect subjective speech perception. In contrast, increased alpha power appears to result from cognitive resource allocation triggered by an interaural mismatch, rather than from conscious recruitment of cognitive resources. Nevertheless, in the paradigm employed in the current study parietal alpha power was a more sensitive measure of the temporal mismatch compared to the pupillometry measure. Combining alpha power with neural and behavioral metrics provides a more comprehensive understanding of the cognitive effort involved when listening with an interaural temporal mismatch.

## 6. CONCLUSION

The current study investigated the effect of cortical temporal mismatch compensation on behavioral speech understanding, pupillometry, CAEPs, selective attention decoding and parietal alpha power in bimodal CI users. To assess the sensitivity of the employed metrics, the study included a listening condition with a significant temporal mismatch of 50ms. This was compared to a condition where the mismatch was compensated (Comp) and the clinical condition (Clin). Key findings of the current study include:

- Speech understanding measures show that despite the presence of a large temporal mismatch, speech understanding remains unchanged across listening conditions.

- In CAEPs, the compensated condition exhibited significantly higher N1P2 amplitudes compared to the temporally mismatched conditions, indicating enhanced neural integration. This finding is further supported by an increase in PLV observed in 8 out of 10 participants when the interaural mismatch was temporally compensated, although this effect was not statistically significant
- With a large interaural temporal mismatch of 50 ms, the morphology of the TRF of selective attention decoding was disrupted, and the N1P2_TRF amplitude significantly decreased, despite unchanged speech understanding. In contrast, the difference between the Clin and Comp listening conditions was less pronounced.
- While pupillometry revealed no differences between listening conditions and showed a correlation to speech understanding for one of the conditions, parietal alpha power significantly increased with a 50ms temporal mismatch, despite unchanged speech understanding performance. No difference in alpha power between the Clin and Comp conditions was observed.

In general, interaural temporal mismatch compensation significantly increases CAEP responses. The lack of significant differences in PLV, N1P2_TRF, and alpha power between the Clin and Comp listening conditions may be attributed to the fact that compensation was only applied in the temporal domain. Therefore, further research is recommended to explore strategies that address both temporal and spectral domain compensation simultaneously.

The primary finding of this study is that neural measurements, such as CAEPs, selective attention decoding, and alpha power, are more sensitive indicators of the effects of interaural mismatches in bimodal CI users. To gain a comprehensive understanding of the cortical mechanisms underlying temporal mismatch and compensation, these neural measures should be used in conjunction with behavioral assessments.

**ACKNOWLEDGMENT**

The authors would like to thank Manuel Segovia-Martinez and Yue Zhang for their support with this study. The authors would also like to thank all subjects who participated in the study.

This work is a part of the project that received funding from the European Research Council (ERC) under the European Union's Horizon-ERC programme (Grant agreement READIHEAR No. 101044753. PI: Waldo Nogueira). Part of these is the part of the project BiMoFuse: Binaural Fusion between Electric and Acoustic Stimulation in Bimodal CI Subjects (ID: 20-1588, PI: Waldo Nogueira) funded by William Demant Foundation. This work was also supported by Deutsche Forschungsgemeinschaft (DFG, German Research Foundation) cluster of excellence "Hearing4all" EXC 2177/1.

**REFERENCE**

(1) https://github.com/sccn/labstreaminglayer

Alain, C. & Arnott ,SR. (2000) Selectively attending to auditory objects. *Front. Biosci*. pp. 202-212

Anderson Gosselin, P., Gagné, JP. (2011) Older adults expend more listening effort than young adults recognizing speech in noise. *J Speech Lang Hear Res.* 54(3):944-58. doi: 10.1044/1092-4388(2010/10-0069).

Baltzel, L., Srinivasan R., Richards, V. (2017) The effect of prior knowledge and intelligibility on the cortical entrainment response to speech. Journal of Neurophysiology 118:6, 3144-3151. https://doi.org/10.1152/jn.00023.2017

Beatty, J., & Lucero-Wagoner, B. (2000). The pupillary system. In J. T. Cacioppo, L. G. Tassinary, & G. G. Berntson (Eds.), *Handbook of psychophysiology* (2nd ed., pp. 142–162). Cambridge University Press.

Belouchrani, A., Abed-Meraim, K., Cardoso, J.-F., Moulines, E. (1993) Second-order blind separation of temporally correlated sources, in *Proc. Int. Conf. on Digital Sig. Proc.*, (Cyprus), pp. 346--351, 1993.

Bizley, JK. & Cohen, YE. (2014) The what, where and how of auditory-object perception. *Nat Rev Neurosci.* 14(10):693-707. doi: 10.1038/nrn3565.

Broadbent, DE. (1958). Perception and communication. Pergamon Press. Doi:

Burg, EA., Thakkar, TD., Litovsky, RY. (2022) Interaural speech asymmetry predicts bilateral speech intelligibility but not listening effort in adults with bilateral cochlear implants. *Front Neurosci.* 7;16:1038856. doi: 10.3389/fnins.2022.1038856.

Carlyon, RP. (2004) How the brain separates sounds. *Trends Cogn.* Sci., 8 (2004), pp. 465-471

Ching, T., van Wanrooy, E., and Dillon, H. (2007). Binaural-bimodal fitting or bilateral implantation for managing severe to profound deafness: a review. *Trends Amplif.* 11, 161–192. doi: 10.1177/1084713807304357.

Crew, J., Galvin, J. r., Landsberger, D., and Fu, Q. (2015). Contributions of electric and acoustic hearing to bimodal speech and music perception. *PLoS ONE* 10, 0120279. doi: 10.1371/journal.pone.0120279

Crosse, M. J., Di Liberto, G. M., Bednar, A., & Lalor, E. C. (2016). The multivariate temporal response function (mTRF) toolbox: A MATLAB toolbox for relating neural signals to continuous stimuli. *Frontiers in Human Neuroscience, 10,* Article 604. https://doi.org/10.3389/fnhum.2016.00604

Darwin, CJ. & Carlyon, RP. (1995) Auditory grouping B.C.J. Moore (Ed.), *Hearing, Academic Press*, Orlando, FL, pp. 387-424.

Decruy, L., Lesenfants, D., Vanthornhout, J., Francart, T. (2020) Top-down modulation of neural envelope tracking: The interplay with behavioral, self-report and neural measures of listening effort. *Eur J Neurosci*. 52(5):3375-3393. doi: 10.1111/ejn.14753.

Delorme, A. & Makeig, S. (2004) EEGLAB: an open-source toolbox for analysis of single-trial EEG dynamics, *Journal of Neuroscience Methods* 134:9-21.

Desimone, R., Duncan, J.(1995) Neural mechanisms of selective visual attention. *Annu Rev Neurosci*. 18:193–222. doi: 10.1146/annurev.ne.18.030195.001205.

Devocht, E., Janssen, A., Chalupper, J., Stokroos, R., and George, E. (2017). The benefits of bimodal aiding on extended dimensions of speech perception: intelligibility, listening effort, and sound quality. *Trends Hear*. 21, 1–20. doi: 10.1177/2331216517727900.

Dieudonné, B., and Francart, T. (2020). Speech understanding with bimodal stimulation is determined by monaural signal to noise ratios: No binaural cue processing involved. *Ear. Hear.* 41, 1158–1171. doi: 10.1097/AUD.0000000000000834

Dieudonné, B., Decruy, L., Vanthornhout, J. (2024) Neural tracking of the speech envelope predicts binaural unmasking. *Eur J Neurosci*. doi: 10.1111/ejn.16638.

Dimitrijevic, A., Smith, M.L., Kadis, D.S. et al. (2019) Neural indices of listening effort in noisy environments. *Sci Rep* 9, 11278. https://doi.org/10.1038/s41598-019-47643-1.

Ding, N. & Simon, JZ. (2012) Neural coding of continuous speech in auditory cortex during monaural and dichotic listening. *J Neurophysiol*. 107:78–89. doi: 10.1152/jn.00297.2011.

Dolhopiatenko, H. and Nogueira, W., (2023). Selective attention decoding in bimodal cochlear implant users. *Frontiers in Neuroscience*, 16, doi: 10.3389/fnins.2022.1057605.

Dolhopiatenko, H., Segovia-Martinez M., Nogueira W. (2024) The temporal mismatch across listening sides affects cortical auditory evoked responses in normal hearing listeners and cochlear implant users with contralateral acoustic hearing. *Hearing Research* 451, doi: https://doi.org/10.1016/j.heares.2024.109088.

Dorman, M., Gifford, R., Spahr, A., and McKarns, S. (2008). The benefits of combining acoustic and electric stimulation for the recognition of speech, voice and melodies. *Audiol. Neurootol.* 13, 105–112. doi: 10.1159/000111782.

Elhilali, M., Ma, L., Micheyl, C., Oxenham, AJ., Shamma, SA.. (2009) Temporal coherence in the perceptual organization and cortical representation of auditory scenes. *Neuron*. 61:317–329. doi: 10.1016/j.neuron.2008.12.005.

Grenzebach, J. & Romanus, E. (2022) Quantifying the Effect of Noise on Cognitive Processes: A Review of Psychophysiological Correlates of Workload. *Noise Health*. 24(115):199-214. doi: 10.4103/nah.nah_34_22

Haas, H. (1951). On the influence of a single echo on the intelligibility of speech, *Acustica* 1, 48–58

Hochmair-Desoyer, I., Schulz, E., Moser, L., Schmidt, M. (1997) The HSM sentence test as a tool for evaluating the speech understanding in noise of cochlear implant users. *Am J Otol*.18(6 Suppl):S83. PMID: 9391610.

Huang, H., Ricketts, TA., Hornsby, BWY., Picou, EM. (2022) Effects of Critical Distance and Reverberation on Listening Effort in Adults. *J Speech Lang Hear Res*. 65(12):4837-4851. doi: 10.1044/2022_JSLHR-22-00109. Epub 2022 Nov 9. PMID: 36351258.

Iotzov, I., Parra, LC. (2019) EEG can predict speech intelligibility. *J Neural Eng*. 16(3):036008. doi: 10.1088/1741-2552/ab07fe.

Just, MA., Carpenter, PA., and Miyake, A. (2003). Neuroindices of cognitiveworkload: neuroimaging, pupillometric and event-related potential studies of brainwork. *Theor. Issues Ergon. Sci.* 4, 56–88. doi: 10.1080/14639220210159735

Kahneman, D. & Beatty, J. (1966) Pupil diameter and load on memory. *Science.* 154(3756):1583-5. doi: 10.1126/science.154.3756.1583. PMID: 5924930.

Kastner, S., Ungerleider, LG. (2000) Mechanisms of visual attention in the human cortex. *Annu Rev Neurosci.* 23:315–341. doi: 10.1146/annurev.neuro.23.1.315.

Koelewijn, T., Zekveld, A., Festen, J., Rönnberg, J., Kramer, S. (2012) Processing load induced by informational masking is related to linguistic abilities. *Int J Otolaryngol,* 865731. doi: 10.1155/2012/865731.

Kong, Y., Stickney, G., and Zeng, F.-G. (2005). Speech and melody recognition in binaurally combined acoustic and electric hearing. *J. Acoust. Soc. Am.* 3, 1351–1361. doi: 10.1121/1.1857526

Kothe, C., Shirazi, SY., Stenner, T., Medine, D., Boulay, C., Grivich, MI., Mullen, T., Delorme, A., Makeig, S. (2024) The Lab Streaming Layer for Synchronized Multimodal Recording. *bioRxiv [Preprint].* doi: 10.1101/2024.02.13.580071.

Krüger, B., Büchner, A., Nogueira, W. (2022) Phantom stimulation for cochlear implant users with residual low-frequency hearing. *Ear. Hear.* 43, 631–645. https://doi.org/10.1097/AUD.0000000000001121

Lesenfants, D., Vanthornhout, J., Verschueren, E., Decruy, L., & Francart, T. (2019). Predicting individual speech intelligibility from the cortical tracking of acoustic- and phonetic-level speech representations. *Hearing Research*, 380, 1-9. https://doi.org/10.1016/j.heares.2019.05.006

Litovsky, R., Johnstone, P., and Godar, S. (2006). Benefits of bilateral cochlear implants and/or hearing aids in children. *Int. J. Audiol.* 45, 78–91. doi: 10.1080/14992020600782956.

MacIntyre, AD., Carlyon, RP., Goehring, T. (2024) Neural Decoding of the Speech Envelope: Effects of Intelligibility and Spectral Degradation. *Trends in Hearing.* 28. doi:10.1177/23312165241266316

McMahon, C., Boisvert, I., de Lissa, P., Granger, L., Ibrahim, R., Lo, C., Miles, K., Graham, P. (2016) Monitoring Alpha Oscillations and Pupil Dilation across a Performance-Intensity Function. *Front Psychol.* 7:745. doi: 10.3389/fpsyg.2016.00745.

Mesgarani, N., Chang, EF. (2012) Selective cortical representation of attended speaker in multi-talker speech perception. *Nature.* 485:233–236. doi: 10.1038/nature11020.

Miles, K., McMahon, C., Boisvert, I., Ibrahim, R., deLissa, P., Graham, P., & Lyxell, B. (2017). Objective assessment of listening effort: Coregistration of pupillometry and EEG. *Trends in Hearing*, 21, 1-13. https://doi.org/10.1177/2331216517706396

Mok, M., Grayden, D., Dowell, R., and Lawrence, D. (2006). Speech perception for adults who use hearing aids in conjunction with cochlear implants in opposite ears. *J. Rehabil. Res. Dev.* 49, 338–351. doi: 10.1044/1092-4388(2006/027).

Nábělek, AK. & Robinette, L. (1978) Influence of the precedence effect on word identification by normally hearing and hearing-impaired subjects. *Journal of the Acoustical Society of America* 63(1): 187–194. DOI 10.1121/1.381711.

Nogueira, W., Dolhopiatenko, H. (2022) Predicting speech intelligibility from a selective attention decoding paradigm in cochlear implant users. *J Neural Eng*. 19(2). doi: 10.1088/1741-2552/ac599f.

Nogueira, W., Cosatti, G., Schierholz, I., Egger, M., Mirkovic, B. and Büchner, A. (2020) Toward Decoding Selective Attention From Single-Trial EEG Data in Cochlear Implant Users. *IEEE Transactions on Biomedical Engineering*, vol. 67, no. 1, pp. 38-49, doi: 10.1109/TBME.2019.2907638.

Näätänen, R. & Picton, T. (1987) The N1 wave of the human electric and magnetic response to sound: a review and an analysis of the component structure. *Psychophysiology*, 24, pp. 375-425.

Obleser, J., Wostmann, M., Hellbernd, N., Wilsch, A., & Maess, B. (2012). Adverse listening conditions and memory load drive a common alpha oscillatory network. *Journal of Neuroscience*, 32, 12376-12383. https://doi.org/10.1523/JNEUROSCI.4908-11.2012

Pichora-Fuller, MK., Kramer, SE., Eckert, MA., Edwards, B., Hornsby, BW., Humes, LE., Lemke, U., Lunner, T., Matthen, M., Mackersie, CL., Naylor, G., Phillips, NA., Richter, M., Rudner, M., Sommers, MS., Tremblay, KL., Wingfield, A. (2016) Hearing Impairment and Cognitive Energy: The Framework for Understanding Effortful Listening (FUEL). *Ear Hear*. Suppl 1:5S-27S. doi: 10.1097/AUD.0000000000000312.

Pieper, S., Hamze, N., Brill, S., Hochmuth, S., Exter, M., Polak, M., Radeloff, A., Buschermöhle, M., Dietz, M. (2022) Considerations for Fitting Cochlear Implants Bimodally and to the Single-Sided Deaf. *Trends Hear.* 26:23312165221108259. doi: 10.1177/23312165221108259.

Polonenko, MJ., Papsin, BC., & Gordon, KA. (2015). The effects of asymmetric hearing on bilateral brainstem function: Findings in children with bimodal (Electric and Acoustic) hearing. *Audiology & Neurotology, 28*(Supp 1), 13–20. https://doi.org/10.1159/000380743.

Reiss, LA., Ito, RA., Eggleston, JL., Wozny, DR. (2014) Abnormal binaural spectral integration in cochlear implant users. *J Assoc Res Otolaryngol*.15(2):235-48. doi: 10.1007/s10162-013-0434-8.

Reiss, LA., Ito, RA., Eggleston, JL., Liao, S., Becker, JJ., Lakin, CE., Warren, FM., McMenomey, SO. (2015) Pitch adaptation patterns in bimodal cochlear implant users: over time and after experience. *Ear Hear*. 36(2):e23-34. doi: 10.1097/AUD.0000000000000114.

Sasaki, T., Yamamoto, K., Iwaki, T., and Kubo, T. (2009). Assessing binaural/bimodal advantages using auditory event-related potentials in subjects with cochlear implants. *Auris Nasus Larynx* 36, 541–546. doi: 10.1016/j.anl.2008.12.001

van der Wel, P., & van Steenbergen, H. (2018). Pupil dilation as an index of effort in cognitive control tasks: a review. *Psychon. Bull. Rev*. 25, 2005–2015. doi: 10.3758/s13423-018-1432-y

Vanthornhout, J., Decruy, L., Wouters, J., Simon, J. Z., & Francart, T. (2018). Speech intelligibility predicted from neural entrainment of the speech envelope. Journal of the *Association for Research in Otolaryngology*, 19, 181. https://doi.org/10.1007/s10162-018-0654-z.

Vanthornhout, J., Decruy, L., Francart, T. (2019) Effect of Task and Attention on Neural Tracking of Speech. *Front Neurosci*. 13:977. doi: 10.3389/fnins.2019.00977.

Vermeire, K., and Van de Heyning, P. (2009). Binaural hearing after cochlear implantation in subjects with unilateral sensorineural deafness and tinnitus. *Audiol. Neurootol*. 14, 163–171. doi: 10.1159/000171478.

Voola, M., Nguyen, AT., Wedekind, A., Marinovic, W., Rajan, G., Tavora-Vieira, D. (2023) A Study of Event-Related Potentials During Monaural and Bilateral Hearing in Single-Sided Deaf Cochlear Implant Users. *Ear and Hearing* 44(4):p 842-853, doi: 10.1097/AUD.0000000000001326.

Warzybok, A., Rennies, J., Brand, T., Doclo, S., & Kollmeier, B. (2013). Effects of spatial and temporal integration of a single early reflection on speech intelligibility. *Journal of the Acoustical Society of America*, 133, 269–282. DOI 10.1121/1.4768880

Wedekind, A., Rajan, G., Van Dun, B., and Távora-Vieira, D. (2020). Restoration of cortical symmetry and binaural function: cortical auditory evoked responses in adult cochlear implant users with single sided deafness. *PLoS ONE* 15, e0227371. doi: 10.1371/journal.pone.0227371

Wess, J., Brungart, D., Bernstein, J. (2017) The effect of interaural mismatches on contralateral unmasking with single-sided vocoders. *Ear and Hearing* 38(3): 374–386. doi:10.1097/AUD.0000000000000374.

Winn, MB., Edwards, JR., Litovsky, RY. (2015) The Impact of Auditory Spectral Resolution on Listening Effort Revealed by Pupil Dilation. *Ear Hear*. 36(4):e153-65. doi: 10.1097/AUD.0000000000000145.

Winn, MB., Wendt, D., Koelewijn, T., Kuchinsky, SE. (2018) Best Practices and Advice for Using Pupillometry to Measure Listening Effort: An Introduction for Those Who Want to Get Started. *Trends Hear*. 22:2331216518800869. doi: 10.1177/2331216518800869.

Wöstmann, M., Herrmann, B., Wilsch, A., Obleser, J. (2015) Neural alpha dynamics in younger and older listeners reflect acoustic challenges and predictive benefits. *J Neurosci*. 35(4):1458-67. doi: 10.1523/JNEUROSCI.3250-14.2015.

Yoon, Y., Shin, Y., Gho, J., and Fu, Q. (2015). Bimodal benefit depends on the performance difference between a cochlear implant and a hearing aid. *Cochlear Implants Int*. 16, 159–167. doi: 10.1179/1754762814Y.0000000101.

Zekveld, AA., Kramer, SE., Heslenfels DJ., Versfeld, NJ., Vriend C. (2024) Hearing Impairment: Reduced Pupil Dilation Response and Frontal Activation During Degraded Speech Perception. *Journal of Speech, Language, and Hearing Research.* Doi: https://doi.org/10.1044/2024_JSLHR-24-00017

Zirn, S., Arndt, S., Aschendorff, A., Wesarg, T. (2015) Interaural stimulation timing in single sided deaf cochlear implant users. *Hearing Research* 328: 148–156. doi:10.1016/j.heares.2015.08.010.

Zirn, S., Angermeier, J., Arndt, S., Aschendorff, A., Wesarg, T. (2019) Reducing the Device Delay Mismatch Can Improve Sound Localization in Bimodal Cochlear Implant/Hearing-Aid Users. *Trends Hear*. 23:2331216519843876. doi: 10.1177/2331216519843876.

## 7. APPENDIX

Figure A1 presents the results of the listening efforts questionnaire. After each list of the presented sentences participants were asked to rate from 1 (not effortful) to 7 (extremely effortful) how effortful to listen to the presented material was.

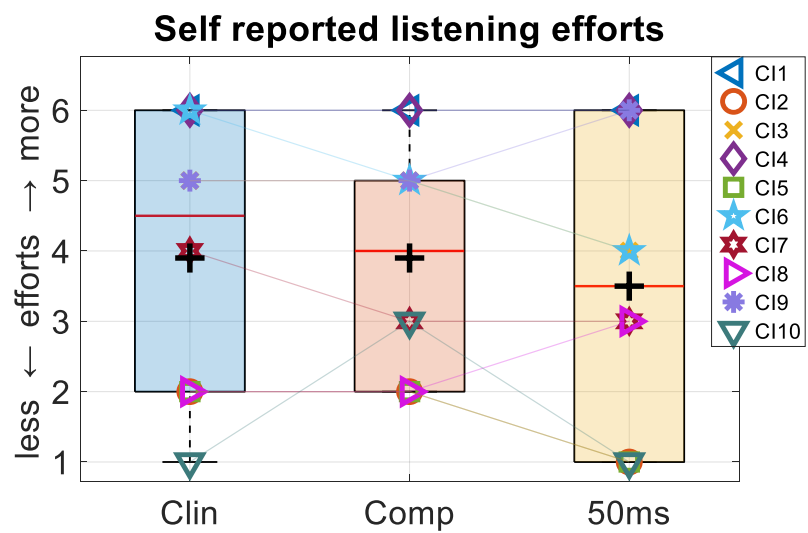

Figure A1. Self-reported listening efforts. After each list of sentences during HSM speech understanding task, participant were asked to rate the listening efforts ranging from 1 – not effortful to 7 – extremely effortful.